\documentclass[sigplan,nonacm]{acmart}
\label{key}
\usepackage{rtospro}
\usepackage{etoolbox}
\usepackage{textcomp}
\usepackage{circledsteps}
\usepackage{subcaption}
\usepackage{pifont}
\usepackage{multirow}

\makeatletter

\date{}
\title{Enabling Deterministic User-Level Interrupts \\in Real-Time Processors via Hardware Extension}

\newtoggle{fullver}
\togglefalse{fullver}

\newtoggle{revision}
\togglefalse{revision}

\usepackage{xcolor}

\iftoggle{revision}{

}{

}

\begin{document}

\author{Hongbin Yang}
\affiliation{%
  \institution{Shandong University}
  \city{Qingdao}
  \country{China}}
\email{hongbinyang@mail.sdu.edu.cn}
 
\author{Huanle Zhang}
\affiliation{%
  \institution{Shandong University}
  \city{Qingdao}
  \country{China}}
\email{dtczhang@sdu.edu.cn}
 
\author{Runyu Pan}
\affiliation{%
  \institution{Shandong University}
  \city{Qingdao}
  \country{China}}
\email{rypan@sdu.edu.cn}


\begin{abstract}
The growing complexity of real-time embedded systems demands strong isolation of software components into separate protection domains to reduce attack surfaces and limit fault propagation.
However, application-supplied device interrupt handlers---even untrusted---have to remain in the kernel to minimize interrupt latency, undermining security and burdening manual certifications.
Current hardware extensions accelerate interrupts only when the target protection domain is scheduled by the kernel; consequently, they are limited to improving average-case performance but not worst-case latency, and do not meet the requirements of critical real-time applications such as autonomous vehicles or robots.

To overcome this limitation, we propose a novel hardware extension that enables direct, deterministic switching to the appropriate protection domain upon user-level interrupt arrival---without kernel intervention---even when that domain is dormant. 
Our hardware extension reduces worst-case latency by more than 50$\times$ with a 19\% increase in core area (2\% of total die area) and 4.1\% increase in dynamic power.
To the best of our knowledge, this is the first integrated mechanism to guarantee user-level interrupt delivery with a nanosecond-scale yet bounded worst-case latency.

\end{abstract}

\maketitle

\section{Introduction}
\label{s:intro}
In modern Cyber-Physical Systems (CPS) such as autonomous vehicles and medical devices, real-time processors execute increasingly complex binaries, comprising software components from heterogeneous sources. 
This software diversity arises from code reuse, multi-stakeholder development, and cost-saving consolidation, deviating significantly from the traditional paradigm where a single team maintains full-stack control over the codebase.

When such diverse components coexist within an unisolated address space, faults or compromises can act as {\em trampolines}, enabling system-wide hijacking.
To mitigate this risk and safeguard these systems, various component isolation strategies have been proposed~\cite{amit17tockos,dejon2023pip,khan2023low,kubica2024mubpf}.
Of these approaches, the microkernel architecture has attracted significant attention due to its compatibility with existing programming paradigms, ease of adoption, and tight security guarantees.
It runs each component as a process, while a privileged microkernel manages scheduling, memory allocation, and device access.
Within this environment, components are isolated
\begin{inparaenum}[(1)]
\item {\em spatially} so that they only access a specific set of memory and peripherals that are explicitly allocated to them, and
\item {\em temporally} so that they cannot exceed their execution budget and interfere with the timely execution of other components.
\end{inparaenum}

However, a critical challenge arises with latency-sensitive real-time interrupt handling.
Ideally, interrupt handlers belonging to untrusted components should run at the user-level to stay spatially and temporally confined.
However, forwarding interrupts from the kernel to user processes inflates latencies by {\em an order of magnitude}, a latency overhead that precludes the sensitive actuator control tasks where deterministic response times are non-negotiable; for example, in closed-loop control, high jitter can cause large control errors or even system failure.

Recent hardware extensions like Intel's UIPI~\cite{intel2025swmanualv3} and RISC-V's N extension~\cite{waterman2025rvpriv} deliver interrupts directly to user-level, but they are {\em opportunistic}.
These mechanisms rely on the target process being scheduled to take the {\em fast path}.
When the target process is dormant, the kernel must intervene to execute domain switching, introducing a {\em slow path}.
This behavior produces an unpredictable bimodal latency distribution that improves only {\em average-case} throughput and fails to enhance {\em worst-case} responsiveness.
To make things worse, when the background runs most of the time, frequent foreground activations always take the slow path, wasting cycles on context switching and reducing average-case throughput.

Consequently, developers are forced down a {\em vicious waterfall}: to obtain tightly bounded worst-case latencies, application interrupt handlers must be placed into the kernel; and to prevent them from launching attacks, laborious certifications of the kernel-resident codebase are required, effectively precluding the use of uncertified binary applications.

\begin{figure}[t]
  \centering
\includegraphics[width=\linewidth,keepaspectratio, trim = 0cm 0.18cm 0cm 0cm, clip]{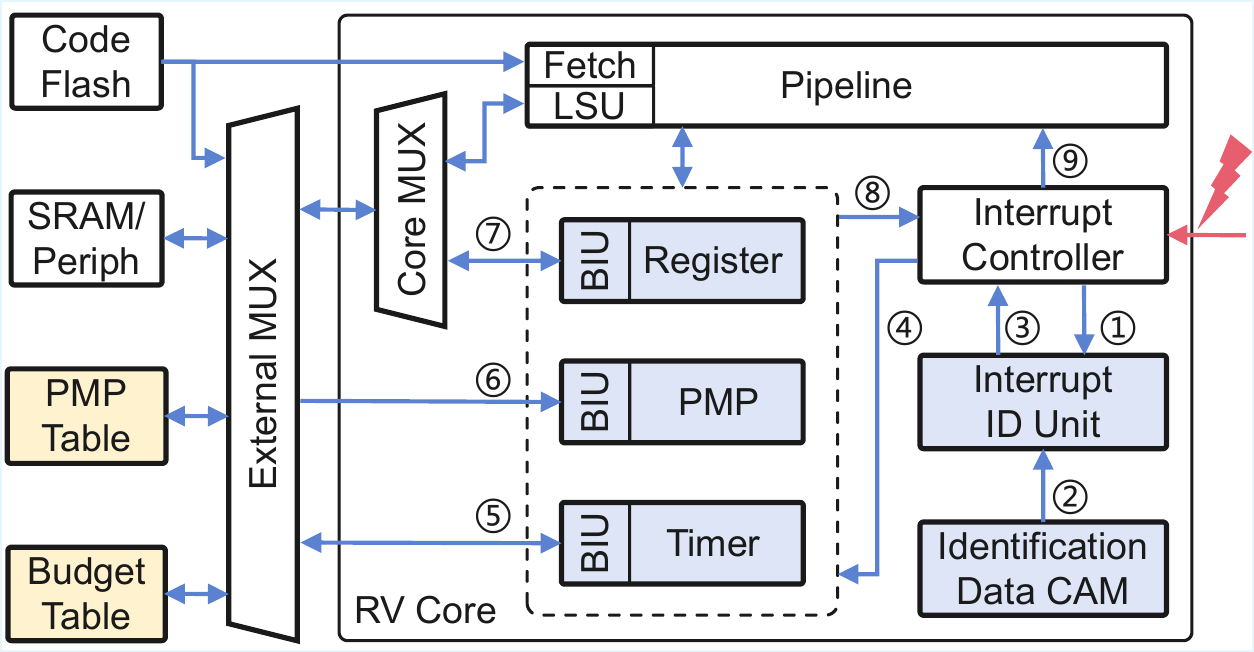}
  \caption{\small Proposed   extension (V5 variant; for detailed variant description, see \S\ref{ss:t2}) block diagram and data flow. The extra TCM blocks are in yellow, and the intra-core extension are in blue.}
  \label{fig:arch}
\end{figure}

To overcome this limitation, this paper proposes a novel hardware extension shown in Figure~\ref{fig:arch} for deterministic user-level interrupt handling, which {\em satisfies real-time determinism requirements without compromising security}.
The extension
\begin{inparaenum}[(1)]
\item enforces secure execution of user-level interrupt handlers via temporal and spatial confinement,
\item enables deterministic user-level interrupt delivery even when the target process is not scheduled, and
\item achieves low delivery latency comparable to that of kernel-level interrupts.
\end{inparaenum}

\head{Contributions.} 
This work presents a hardware extension that enables deterministic user-level interrupt delivery and secure confinement of handler execution without {\em any} software mediation.
Its key contributions are as follows:
\begin{itemize}
\item {\bf we explore the hardware design space}
(\S\ref{s:design}) for deterministic user-level interrupt delivery with spatial and temporal execution confinement, reducing worst-case user-level interrupt latency by more than {\em 50$\times$};
\item {\bf we propose a system-level abstraction}
(\S\ref{s:design}) compatible with existing microkernel operating systems to efficiently leverage the proposed hardware mechanism;
\item {\bf we implement select variants} (\S\ref{s:impl}) on the RISC-V architecture using a simulated 45nm process, discussing the Performance, Power, and Area (PPA) trade-offs, demonstrating a minimum power and area overhead of only 4.2\% and 19\%, respectively;
\item {\bf we evaluate key implementations} (\S\ref{s:eval}) with Field-Programmable Gate Array (FPGA) prototyping, demonstrating up to {\em 60$\times$} jitter reduction on latency-sensitive tasks and {\em 11$\times$} CPU load reduction on high-speed communication tasks, both of which are common in the industry.
\end{itemize}

To the best of our knowledge, this work presents the first user-level interrupt delivery mechanism with strong real-time determinism, enabling seamless consolidation of multiple real-time controllers without sacrificing interrupt latency.

\section{Background and Related Work}
\label{s:related}
\subsection{User-level Interrupt Delivery}
The emergence of kernel-bypassing technologies has transformed high-performance cloud computing.
User-level scheduling~\cite{gadepalli2020slite}, networking~\cite{dpdk} and storage stacks~\cite{spdk} shift functions traditionally handled by the kernel to the user-space, eliminating unnecessary processor state transitions, leading to improved system efficiency.
A recent addition is user-level interrupts, which delivers interrupts directly to the user-space without kernel intervention.

The most notable implementations are the Intel's UIPI extension~\cite{intel2025swmanualv3} and RISC-V's N extension~\cite{waterman2025rvpriv,rv2025next}, with the former commercially available and the latter in unratified draft.
Intel’s UIPI extension enables user-level interrupt delivery by having the kernel configure a User Interrupt Target Table (UITT).
A sender thread issues a SENDUIPI instruction with a UITT offset, and the hardware resolves the target thread, triggering an IPI via the APIC to activate the handler.
While designed to reduce cross-core IPC overhead, the Sapphire Rapids implementation restricts senders to threads, excluding devices, which limits its relevance for CPS.
In contrast, the S, H, and N extensions of RISC-V introduce a unified interrupt delegation model, allowing per-mode delegation registers to route interrupts to less privileged modes regardless of whether the sender is a thread or a device.

Some research have accumulated around user-level interrupts, encompassing applications like accelerated user-level scheduling~\cite{li2024libpreemptible} and tail latency optimization~\cite{jia2024skyloft}.
Additionally, one study focused on optimizing hardware for efficient interrupt delivery~\cite{aydogmus2025xui}.
In a broader context, user-level interrupts can be seen as a subtype of {\em deprivileging interrupts}, delivering interrupts to a lower privilege level instead of a higher one.
This definition shares similarities with virtualization extensions like Intel's VT-d posted interrupts, which have been leveraged to achieve performance boosts similar to those found in user-level interrupts~\cite{kaffes2019shinjuku}.
However, existing research on user-level interrupts prioritizes average performance over deterministic latency, as the OS kernel still manages scheduling and protection domain switching.

\subsection{Hardware Thread Schedulers}
The real-time community have long been using hardware as schedulers to bring down scheduling latencies \eg~\cite{stanischewski1993fastchart,nakano1995hardware,daleby2002rtumanual,morton2004hardware}.
In this scheme, the hardware may handle scheduling decisions, register context switching, or both.

When scheduling algorithms are implemented in hardware, they tend to have a low and deterministic latency, especially when accelerating complex algorithms like Earliest Deadline First (EDF)~\cite{tang2015hardware}, or the number of processor cores are large~\cite{derafshi2020hrhs}.
However, pure hardware schedulers require customized hardware, and fixed task parameters, which can be limiting.
To overcome this, some research use prioritization logic of interrupt controllers~\cite{hofer2009sloth,daniel04safersloth} to perform scheduling.

When register context switching is implemented in hardware, software saving or restoring sequences are bypassed, resulting in lower latency~\cite{rafla2011hardware,scheck2026coexp}.
Two approaches exist:
\begin{inparaenum}[(1)]
\item automatic register pushing and popping on exception entries and returns~\cite{yiu2009definitive,balas2024cv32rt}, and
\item providing additional partial~\cite{armv5} or full~\cite{steiner20058051,grunewald1996towards,dodiu2012custom,qingkev4_trm,nurmi2025efficient} register banks for different processing contexts.
\end{inparaenum}
Most register context switch mechanisms don't involve process switches, except for x86's outdated task gate~\cite{intel2025swmanualv3}.
However, this mechanism incurs large overhead, is poorly optimized at the hardware level, and can introduce unpredictable memory access latencies due to TLB flushing, all of which probably led to its removal in 64-bit mode.

To sum up, we are not aware of any hardware mechanism that performs all of register context switching, budget-enforcing scheduling and access permission switching in a swift fashion.
It is noteworthy that~\cite{pan2022sbis} and~\cite{pinto19voila} leverage ARM TrustZone-M to accelerate user-level interrupt delivery by assigning them to a dedicated TrustZone world, however these approaches are working against the hardware's intention and thus is limited to only one process.

\subsection{Real-time Processor Microkernels}
Modern real-time processors feature privilege levels and Physical Memory Protection (PMP), enabling microkernel deployment.
In contrast to general-purpose Memory Management Unit (MMU), PMP omit hardware address translation and TLB structures, using software-managed, segment-based physical address protection with deterministic access latency.
Such choice is deliberate as TLB misses introduce unpredictable latency, which may violate hard real-time constraints.
However, this restricts these systems to Single Address Space Operating Systems (SASOSes).

Many notable SASOS microkernels~\cite{huangf9micro,pan18mpu,sundar2018impl,pan2025fvm} have leveraged PMP as isolation facilities.
Besides, various other works also explored PMP-based isolation approaches~\cite{daniel04safersloth,amit17tockos,sa2021first,oliveira2022utango} with similar takes.
However, these works all suffer from significant execution overhead that arise from processor mode switches, with interrupt latencies reaching hundreds of clock cycles or more.
While this is acceptable for less demanding cases, such delays exceed the tolerance for scenarios where tens of clock cycles are considered the limit.

\iftoggle{fullver}{
\subsection{Non-microkernel Isolation Surrogates}
In addition to microkernel-based isolation approaches, various surrogates can operate within a single address space.
Some approaches rely solely on software and can go without hardware isolation facilities of any sort.
This includes
\begin{inparaenum}[(1)]
\item Hopter~\cite{ma2025hopter}, ARMor~\cite{zhao11armor}, CRT-C~\cite{khan2023low} and Minion~\cite{kim18minion} that leverage bounds checking or compiler-based verifications, and
\item eBPF~\cite{zandberg2022femto,kubica2024mubpf}, Java~\cite{microej}, Python~\cite{micropython}, Javascript~\cite{duktape,mjs} and even WebAssembly~\cite{singh19warduino,peach20ewasm} that rely on managed execution environments.
\end{inparaenum}
Some approaches resort to a combination of software and hardware, such as TockOS~\cite{amit17tockos}, uXOM~\cite{kwon19uxom} and Pip-MPU~\cite{dejon2023pip}, for efficiency.
Outside the microcontroller domain, researchers have explored isolation facilities that combine software with Intel MPK, ARM memory tagging, or hardware capabilities~\cite{vahldiek2019erim,hedayati2019hodor,li2020iso,sartakov2021cubicleos,kuenzer2021unikraft,voulimeneas2022you,almatary2022compartos,dinh2023capacity,shao2025janus}.
Notably, these efforts also focus on a single address space such as a process or unikernel.

While these approaches have their merits, they often come with significant limitations.
Some require toolchain modifications that are unfeasible in a production setting, while others rely on tedious manual isolation boundary annotations.
Additionally, these approaches often restrict the language selection, neglect availability concerns, assume asymmetric isolation, cannot be linked against third-party binaries or require binary scanning, adding complexity and overhead.
Furthermore, some of them incur high execution overheads and large memory footprints, making them suitable only for resource-rich environments.
}{
We do note that various compiler-, language-, or sandbox-based surrogates \eg~\cite{amit17tockos,peach20ewasm,zandberg2022femto,kubica2024mubpf,ma2025hopter} can also achieve isolation within a single address space.
However, they can require impractical toolchain changes, tedious annotations, or restrictive language choices, and typically precludes linking with {\em unsafe} binaries.
Some of them incur high runtime overhead and large memory footprints, limiting their use to resource-rich environments.
}

\subsection{RISC-V Real-time Processors}
RISC-V has emerged as a prominent architecture in the embedded domain, driven by its flexibility and open-source ecosystem.
This openness has spurred widespread adoption, with numerous vendors (e.g. WCH, GigaDevice) and IP providers (e.g. SiFive, Nuclei) offering RISC-V-based solutions.
They typically feature pipelined in-order cores, ranging from hundreds of kilobytes to megabytes of memory, and are often connected to memory via 32-bit or 64-bit AMBA AHB-lite buses.

\subsection{Threat and Task Model}
\head{Threat Model.}
We assume mutually distrusting user-level handlers attempting spatial (illegal access) or temporal (CPU hogging) violations.

\head{Task Model.}
Run-to-completion aperiodic user-level interrupt handlers are hosted on deferrable servers with strict budget constraints, which are enforced by the hardware extension.

\section{Design Space Exploration}
\label{s:design}
The primary objective of this extension is to ensure deterministic user-level interrupt delivery by entirely eliminating software mediation.
It aims to achieve the following goals:
\begin{itemize}[topsep=1pt,itemsep=0pt,parsep=1pt,itemindent=0pt,leftmargin=0.3in]
\item[{\bf G1:}] {\bf Deterministic latency.}
  User-level interrupt handlers must commence execution with a deterministic and low latency upon activation regardless of the currently scheduled process, ensuring system responsiveness.
\item[{\bf G2:}] {\bf Spatial confinement.}
  User-level interrupt handlers must have their memory and device accesses confined to authorized memory regions, protecting system confidentiality and integrity.
\item[{\bf G3:}] {\bf Temporal confinement.}
  User-level interrupt handlers must execute within a rigidly enforced budget, preventing unbounded execution from interfering with other components, safeguarding system availability.
\item[{\bf G4:}] {\bf Minimal hardware overhead.}
  The extension must incur minimal silicon area and power overheads, respecting the Size, Weight, and Power plus Cost (SWaP-C) constraints of modern CPSes.
\item[{\bf G5:}] {\bf Ecosystem compatibility.}
  The extension must incur minimum changes to the bus interface of the original CPU core to ensure applicability across diverse architectures, and provide a companion software abstraction to facilitate integration into existing microkernels.
\end{itemize}

\head{Latency vs. Area vs. Power Trade-off.}
Lower interrupt latency typically comes at the cost of increased hardware complexity and power consumption to support parallel microarchitectural operations.
We aim to characterize the latency–area–power trade-off and strike a balance that meets real-time responsiveness requirements without excessive overheads.
  
\subsection{Real-time Processor Core}
\begin{table}[H]
\scriptsize
\centering
\caption{\small Comparison of typical real-time processors. BP denotes Branch Prediction, interrupt latency is in CPU cycles, CoreMark is per MHz when running from zero wait-state memory. *The datasheet does not contain this data.}
\begin{tabular}{|c|c|c|c|c|}
  \hline Processor    & Stages                  & BP                        & CoreMark       & Interrupt Latency \\
  \hline Cortex-M0    & \multirow{3}{*}{3}      & \multirow{4}{*}{Static}   & 2.33           & 16      \\
  \cline{1-1} \cline{4-5}
  Cortex-M3           &                         &                           & 3.45           & \multirow{2}{*}{12}      \\
  \cline{1-1} \cline{4-4}
  Cortex-M4           &                         &                           & 3.54           &         \\
  \cline{1-2} \cline{4-5}
  Cortex-M23          & 2                       &                           & 2.64           & 15      \\
  \hline Cortex-R4    & 8                       & \multirow{2}{*}{Dynamic}  & 3.47           & 20      \\
  \cline{1-2} \cline{4-5}
  QingKe V4    & 3                       &                           & 2.83           & *       \\
  \hline
\end{tabular}
\label{tbl:otscpu}
\end{table}

To explore the design space, a flexible CPU backbone is needed.
In this light, we aim to design a RISC-V core with a 3-stage pipeline, static branch prediction, a FPU, and a performance around 3 CoreMark/MHz as our baseline core, which is in accordance with most popular processor cores~\cite{arm_cortex_m3_trm} on the market shown in Table~\ref{tbl:otscpu}.

Note that this core is without MMU, superscalar capability, or caches, aligning with area and power performance requirements of mainstream applications.

\subsection{Hardware Extension Synopsis}
The extension allows confining specific user-level interrupt sources to 
\begin{inparaenum}[(1)]
\item a Spatial Protection Domain (SPD) described by a set of Physical Memory Protection (PMP) configurations {\bf (G2)}, and 
\item a Temporal Protection Domain (TPD) enforced by a hardware budget countdown timer {\bf (G3)},
\end{inparaenum}
which will be switched to when the interrupt source is activated, then the whole register set will be saved by the hardware.
The extension also distinguishes user-level interrupt sources from traditional kernel-level interrupt sources, maintaining backward compatibility.
We showcase the operation of the extension's internal mechanisms with a concrete implementation that we call the V5 variant (see \S\ref{ss:t2}).
Its block diagram are shown in Figure~\ref{fig:arch}.

Upon interrupt entry, the Interrupt Controller (IC) calls upon the Interrupt Identification Data Unit (IIDU) to distinguish \Circled{1} whether this is a user-level interrupt.
If it is, IIDU identifies \Circled{2} the address of the PMP configuration and budget value from the Interrupt Identification Data (IID) it manages, which is a Content Addressable Memory (CAM) in this variant, and notifies the IC \Circled{3}.
Subsequently, the IC triggers \Circled{4} parallelizable hardware operations where
\begin{inparaenum}[(1)]
\item the execution budget is loaded \Circled{5}, in this variant, from Tightly Coupled Memory (TCM),
\item the PMP configuration is loaded \Circled{6} from its TCM as well,
\item the registers are saved by switching banks or spilling to memory \Circled{7} then clearing.
\end{inparaenum}

Once completed, a finish signal is sent \Circled{8} to the IC, after which it resumes \Circled{9} the interrupt vector and the timer countdown.
When the interrupt vector executes, if
\begin{inparaenum}[(1)]
\item the PMP detects a out-of-bound access which indicates a spatial violation, or
\item the timer counts down to zero which indicates a temporal violation,
\end{inparaenum}
a user-level interrupt return is forced to maintain strong isolation.

Upon interrupt return, the register context and PMP configuration is restored, with the remaining execution budget written back to the budget table \Circled{5}.
Were spilling occurred during stacking, the register context is restored from memory \Circled{7}.

\subsection{User-level Interrupt Identification}
Identifying which interrupt source should be directed to the user-level and what protection domain it targets requires querying a hardware-managed IID structure managed by the IIDU.
The design must balance lookup latency {\bf (G1)} against silicon area {\bf (G4)}, as this step is on the critical path.

\head{\bf IID location.}
Placing IID in CPU registers ensures deterministic, single-cycle lookup for its target protection domain information, but increases the core area.
Conversely, storing IID in RAM saves logic area but introduces additional bus latencies, undermining {\bf G1}.

\head{\bf IID organization.}
The IID maps interrupt numbers to two protection domain numbers and can be designed in two ways.
A linear table indexed by interrupt number offers simplicity but requires substantial storage proportional to the vector table size. Alternatively, a compressed form—--suitable given that only select interrupts are user-level—--stores explicit interrupt and domain number entries, necessitating the use of CAM for timely lookups {\bf (G4)}.

\begin{table}[htb]
\scriptsize
\centering
\caption{IID design summary. Feasible options are in bold.}
\begin{tabular}{|m{1.3cm}<{\centering}|m{3cm}<{\centering}|m{3cm}<{\centering}|}
  \hline IID                           & In CPU                                                   & In RAM \\
  \hline \rule{0pt}{3.2ex}As table     & \makecell{Infeasible;\\table too big to fit into CPU}    & \makecell{{\bf Less CPU logic}\\{\bf more latency}} \\
  \hline \rule{0pt}{3.2ex}As CAM       & \makecell{{\bf Less latency}\\{\bf more core area}}  & \makecell{Infeasible;\\ CAM needs parallel lookup} \\
  \hline
\end{tabular}
\label{tbl:iiddesign}
\end{table}

\head{\bf Discussion.}
Combining these factors, we dismiss the area-prohibitive CPU-Table approach and the latency-prohibitive RAM-CAM approach, narrowing the design space to two viable options shown in Table~\ref{tbl:iiddesign}.
To {\em maintain compatibility with the RISC-V standard}, we always separate IID from the original interrupt vector table in the linear table variant.

\subsection{PMP Table and Budget Table}
The PMP table stores spatial confinements for each domain {\bf (G2)}, while the budget table enforces temporal isolation {\bf (G3)}. 
Each PMP table entry consists of a set of configurations loaded into the PMP CSRs when switching the spatial protection domain.
Conversely, each budget table entry maintains the number of CPU clocks loaded into the countdown timer when the interrupt executes.
Given their size, both tables must reside in memory, necessitating a trade-off between bus interface complexity and the potential for bus contention.

\head{\bf Tables in main SRAM.}
Given that both tables are consulted sequentially after the IID lookup, they can coexist with the IID in the same SRAM.
However, parallelization of context saving and table accesses creates potential contention on the shared SRAM bus, introducing latency spikes {\bf (G1)}.
On the bright side, this approach requires no additional memory ports and maintains full compatibility with the original core's Bus Interface Unit (BIU) {\bf (G5)}.

\head{\bf Tables in tightly-coupled memory.}
To eliminate bus contention, the PMP and budget table could be placed in one shared or two dedicated SRAM blocks. 
However, this requires one or even two additional memory ports, complicating the bus interface design {\bf (G4)}.
Fortunately, many processors {\em come architecturally with optional instruction or data TCM ports}, which we can take advantage of if there is.

\head{\bf Tables in flash.}
Similar to TCM, many System-on-Chip (SoC) feature a dedicated bus for read-only embedded code flash that offers throughput comparable to SRAM with higher access latency, which can accommodate the tables.
However, the remaining budget must be written back to the budget table upon interrupt return to facilitate the implementation of bandwidth-based servers, and the kernel may periodically replenish these budgets.
Consequently, only the PMP table may reside in flash, while the budget table must be held in RAM.

\begin{table}[htb]
\scriptsize
\centering
\caption{\small Protection table design summary.}
\begin{tabular}{|m{1.4cm}<{\centering}|m{3cm}<{\centering}|m{3cm}<{\centering}|}
  \hline Tables in                 & Advantages                                                     & Disadvantages \\
  \hline \rule{0pt}{3.2ex}Main SRAM   & \makecell{Simplistic design;\\no changes to bus interfaces}    & \makecell{Potential bus conflicts;\\longer latency} \\
  \hline \rule{0pt}{3.2ex}Data TCM    & \makecell{No potential bus conflicts;\\likely less latency}    & \makecell{Requires TCM interface;\\complicates hardware} \\
  \hline \makecell{Code Flash\\(PMP Only)}  & \makecell{Simplistic design;\\no potential bus conflicts}      & \makecell{Flash cannot be modified;\\slightly longer latency} \\
  \hline
\end{tabular}
\label{tbl:pmptdesign}
\end{table}

\head{\bf Discussion.}
All three options shown in Table~\ref{tbl:pmptdesign} are feasible, each with its own trade-offs.
However, we will only explore the first two options in the implementation section, as flash is a degenerated surrogate for TCM for our purposes.

\subsection{Register Context Saving}
To prevent untrusted user-level handlers from peeking the register set of threads or handlers that belong to other processes, the hardware must save the full register set and zeroize it.
Given the size of the register set, this is the {\em primary bottleneck} for latency {\bf (G1)}, and {\em conflicts with other activities should be avoided}.

\head{Register shadow banking.}
In this scheme, a dedicated register bank is reserved for handling interrupts, thereby drastically reducing context switching latency by eliminating the need for register stacking {\bf (G1)}.
However, this approach
\begin{inparaenum}[(1)]
\item significantly increases core area {\bf (G4)}, and
\item does not support nesting unless multiple banks are available.
\end{inparaenum}

\head{Register hybrid spilling.}
To support unbounded nesting without unbounded area, we implement a fallback mechanism.
When shadow banks are exhausted (or no bank at all), the hardware automatically spills the context to memory.
The spilled register context can be written to SRAM—-which may introduce bus conflicts and prolong interrupt latency--or utilize a dedicated TCM port.

\begin{table}[H]
\scriptsize
\centering
\caption{\small Register context design summary.}
\begin{tabular}{|m{1.7cm}<{\centering}|m{2.7cm}<{\centering}|m{3cm}<{\centering}|}
  \hline Register in                                      & Extra banks with spilling                                                                        & Stacking only (no banks)\\
  \hline \rule{0pt}{4.3ex}Main SRAM      & \makecell{Spills less but more area;\\longer stacking latency;\\simple bus interface}  & \makecell{Always spills but less area;\\longer stacking latency;\\simple bus interface} \\
  \hline \rule{0pt}{4.3ex}Dedicated TCM            & \makecell{Spills less but more area;\\shorter stacking latency;\\additional TCM interface}  & \makecell{Always spills but less area;\\shorter stacking latency;\\additional TCM interface} \\
  \hline
\end{tabular}
\label{tbl:ctxdesign}
\end{table}

\head{Discussion.}
We explore all four options listed in Table~\ref{tbl:ctxdesign}.
Since the number of banks primarily affects nesting depth before latency issues arise, we only vary the number of banks when examining its impact on core area.

\subsection{Protection Domain Context Saving}
Since the extension preempts all underlying execution including the kernel, the preempted protection domain information must be saved to remain compatible with the software ecosystem {\bf (G5)}.
This includes
\begin{inparaenum}[(1)]
\item the kernel-managed PMP settings, which reflect the access permissions of the current process, and
\item the current system timer value, which reflects the budget of the current microkernel thread.
\end{inparaenum}
Furthermore, user-level interrupts must be capable of preempting one another.

\head{User-level interrupt being preempted.}
This is a straightforward case: since all PMP settings for user-level interrupts are stored in their respective tables, no additional storage is required; they simply need to be reloaded upon resumption.
Regarding the budget timer, however, there is a subtle distinction: when preemption occurs, the remaining time must be written back to the budget table to account for the time expended, and subsequently reloaded when the interrupt resumes.

\head{Thread or kernel being preempted.}
This presents a complex scenario, as the PMP context at the point of preemption is managed by the kernel.
To address this, we can either
\begin{inparaenum}[(1)]
\item provide an extra set of shadow PMP registers, or
\item spill the PMP registers to memory.
\end{inparaenum}
Additionally, to prevent the execution time of user-level interrupt handlers from being attributed to kernel-managed threads, the system timer must be paused.

\subsection{Ancillary Hardware Mechanisms}
\head{Exception handling.}
Handling exceptions in our user-level interrupts poses a challenge because they preempt kernel execution.
As exception handlers are themselves kernel code, triggering exception handling for user-level interrupts may cause kernel reentrancy.
To address this issue, we let the user-level interrupt handler returns immediately upon any exception, with the cause  recorded in an {\em optional} dedicated Control and Status Register (CSR) which the kernel may inspect asynchronously.
Similarly, if the user-level interrupt interrupt handler exhausts its budget before returning, a forced return occurs.
In the end, as user-level interrupts {\em are part of the application}, is the responsibility of the application developer to ensure exception-free operation of them.

\head{Tail-chaining opportunity.}
The context switch for user-level interrupts is heavier than for kernel-level interrupts, as it entails protection domain switching.
To minimize switches during back-to-back execution, we could employ tail-chaining mechanism~\cite{yiu2009definitive} to elide the previous interrupt’s context restore and the next interrupt’s context save if they target the same spatial protection domain.
Although this improve {\em average-case} performance, we do not implement it since optimizing {\em worst-case} latency is the focus of this work.

\iftoggle{fullver}{
\head{FPU context lazy saving opportunity.}
Real-time cores have long been equipped with FPUs, with recent cores like Cortex-M85 featuring even SIMD extensions, highlighting the growing importance of floating point capabilities.
To optimize stack pushing latency, we can reserve space for FPU registers and only trigger actual saves and restores when FPU instructions are actually used, as such uses are infrequent in interrupt handlers~\cite{yiu2009definitive}.
Since FPU handling falls outside the scope of our paper's focus, we do not implement the FPU, let alone this optimization.
}

\subsection{System-level Software Abstraction}
\begin{table}[h]
\scriptsize
\centering
\caption{\small Core API summary.}
\begin{tabular}{|c|c|}
  \hline System call         & Functionality  \\
  \hline int\_reg()          & Register an user-level interrupt. \\
  \hline int\_del()          & Delete a registered user-level interrupt. \\
  \hline int\_prio()         & Configure user-level interrupt preemption priority. \\
  \hline int\_ena()          & Enable an registered user-level interrupt. \\
  \hline int\_dis()          & Disable an registered user-level interrupt. \\
  \hline
\end{tabular}
\label{tbl:sysapi}
\end{table}

To integrate user-level interrupts into the software ecosystem, we present a system-level APplication Interface (API) shown in Table~\ref{tbl:sysapi} that abstract interrupts as hardware threads.
Hardware manages their execution context, while scheduling co-managed: the hardware enforces preemption policies, and the kernel handles budget replenishment.
In this light, we further introduce hardware budget objects to facilitate this replenishment.

\begin{figure}[h]
  \centering
\includegraphics[width=\linewidth, trim = 0cm 0.08cm 0cm 0cm, clip]{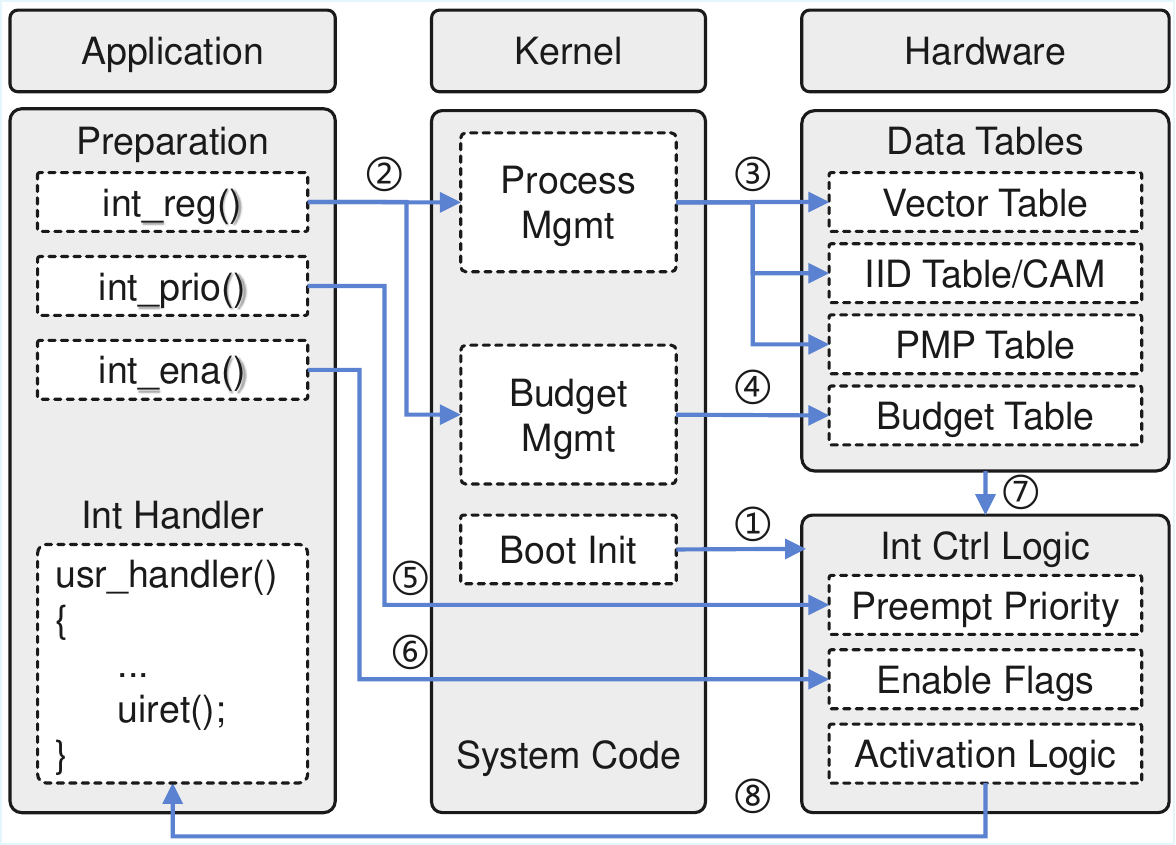}
  \caption{\small User-level interrupt API and software-hardware workflow.}
  \label{fig:api}
\end{figure}

At boot time, the kernel initializes the hardware extensions \Circled{1} to enable user-level interrupt operations for user processes.
To register a user-level interrupt handler, the application first invokes the {\tt int\_reg(int\_id, entry, policy)} \Circled{2}, which passes the interrupt number, entry address and desired scheduling policy to the kernel's process and budget management facilities after due permission checks.
The process management module allocates a free entry in the IID table, registers the handler in the interrupt vector table, and configures the PMP entry to enforce the application's access permissions \Circled{3}.
If a PMP entry is already allocated to the application, the existing entry is reused rather than allocating a new one.
The budget management module programs the execution budget into the corresponding budget table entry \Circled{4}, where the allocation, reuse, and replenishment of an entry is determined by the specified policy.
The application then sets the interrupt priority via {\tt int\_prio(prio)} \Circled{5} and enables the interrupt via {\tt int\_ena()} \Circled{6}.
When the interrupt is triggered, the interrupt control logic loads configuration from the hardware tables \Circled{7} and redirects the pipeline to execute the user-level handler \Circled{8}.

\section{Implementation and Optimization}
\label{s:impl}
\begin{figure*}[t]
  \centering
  \includegraphics[width=\textwidth,trim=0.35cm 0.12cm 0cm 0cm, clip,keepaspectratio]{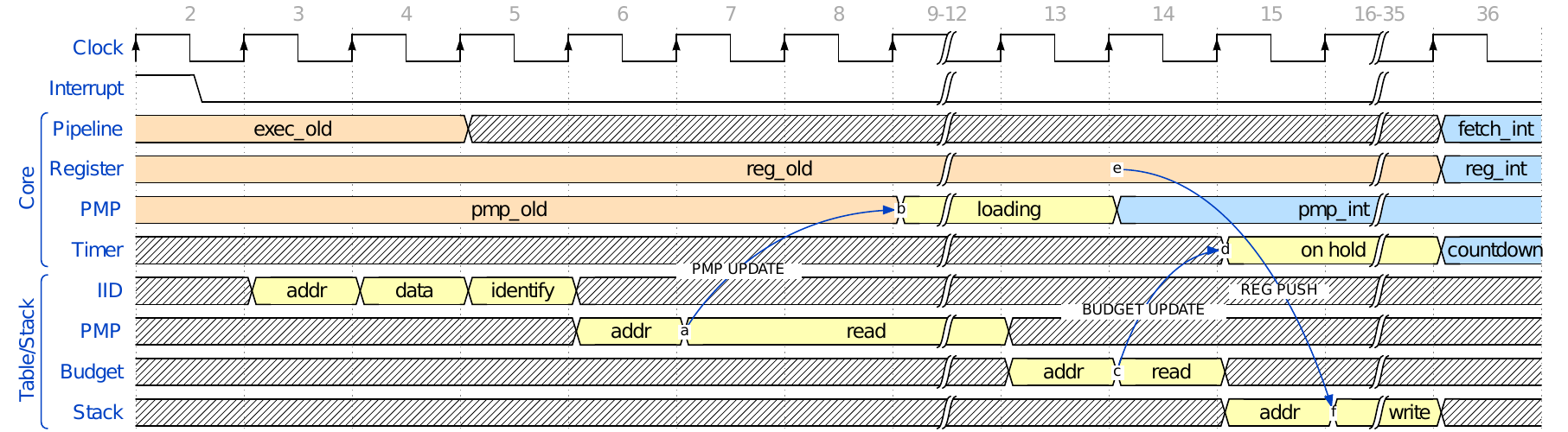}
  \caption{\small Timing diagram of the V1 variant upon user-level interrupt activation, with a latency of 38 cycles (2 more cycles are needed for the first fetched interrupt vector instruction to reach the execute stage). a$\to$b shows PMP table consulting and PMP updating, c$\to$d shows budget table consulting and timer updating, while e$\to$f shows register stacking. The kernel-managed PMP is shadow banked and not shown.}
  \label{fig:c1var}
\end{figure*}

\begin{figure*}[]
  \centering
  \includegraphics[width=\textwidth,trim=0.35cm 0.10cm 0cm 0cm, clip,keepaspectratio]{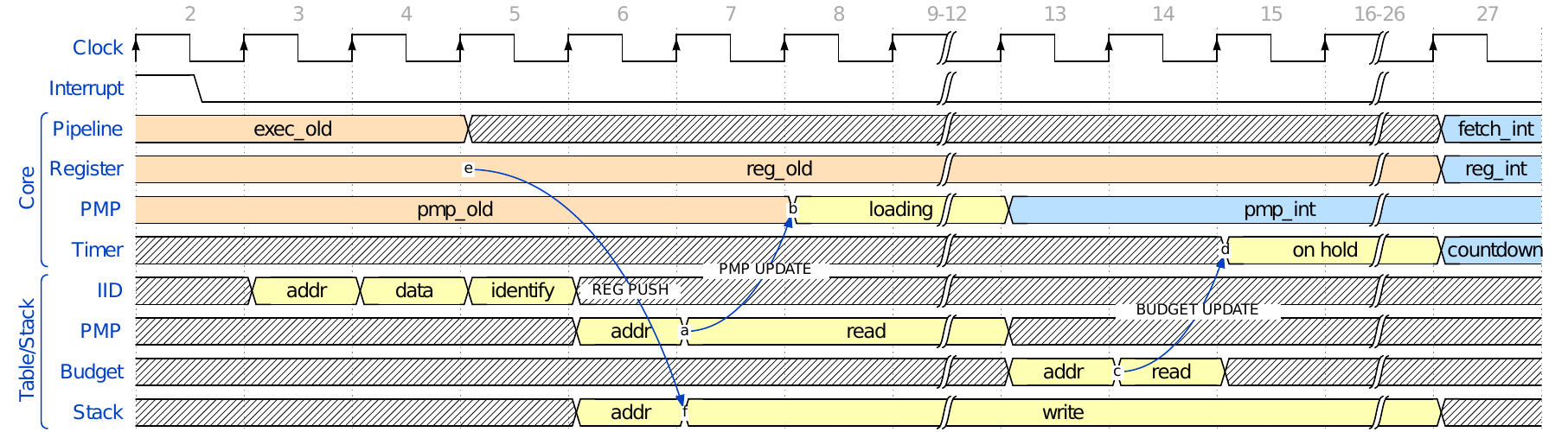}
  \caption{\small Timing diagram of the V2 variant upon user-level interrupt activation, with a latency of 29 cycles. The commentary is the same as that of Figure~\ref{fig:c1var}. The deciding factor here however, is still register context stacking.}
  \label{fig:c2var}
\end{figure*}

\subsection{Implementation Primer}
We synthesize and lay out our RISC-V core with the extension on the Nangate45 platform using OpenROAD~\cite{ajayi2019openroad} to derive its silicon area, and on Xilinx XC7K410T FPGA to obtain runnable instances.
The power metrics are then measured using OpenROAD's built-in flow, simulating a computation of a single-layer convolutional neural network on the processor.
Finally, we use both silicon area, core power and interrupt latency performances to guide our search iteratively.

\subsection{Baseline RISC-V Core}

\begin{table}[H]
\scriptsize
\centering
\caption{\small Characteristics of different processor configurations. CoreMark is per MHz, area is in mm{$^2$}, latency is in CPU cycles, and power is in static (mW) / dynamic ($\mu$W/MHz) format. ``Naked'' is the core logic, ``+PMP'' adds PMP, ``+FPU*'' adds a FPU that is not powered, ``+SBP'' adds static branch prediction, ``+MDU'' adds multiplier and divider, ``Base'' is Naked+PMP+FPU*+SBP+MDU.}
\begin{tabular}{|c|c|c|c|c|}
  \hline Processor        & CoreMark                            & Floorplan Area          & Latency                       & Power \\
  \hline Naked (RV32I)    & \multirow{3}{*}{0.86}               & 0.0252 (27\%)           & \multirow{5}{*}{5}            & 0.527/25.62 \\
  \cline{1-1} \cline{3-3} \cline{5-5}
  Naked+PMP               &                                     & 0.0332 (36\%)           &                               & 0.704/25.68 \\
  \cline{1-1} \cline{3-3} \cline{5-5}
  Naked+FPU*              &                                     & 0.0720 (78\%)           &                               & 1.61/25.62 \\
  \cline{1-3} \cline{5-5}
  Naked+SBP               & 0.90                                & 0.0258 (28\%)           &                               & 0.533/28.12 \\
  \cline{1-3} \cline{5-5}
  Naked+MDU               & 2.66                                & 0.0362 (39\%)           &                               & 0.738/27.36 \\
  \hline {\bf Base }      & {\bf 2.84}                          & {\bf 0.0931 (100\%)}    & {\bf 5}                       & {\bf 2.07/30.48} \\
  \hline Cortex-M3 (DS)   & 3.45                                & 0.0889                  & 12                            & N/A \\
  \hline IBEX (RV32IMC)   & 2.36                                & 0.0472                  & 6                             & N/A \\
  \hline
\end{tabular}
\label{tbl:ourcpu}
\end{table}

To evaluate the extension, we first implement the baseline RISC-V processor supporting the RV32IM instruction set, and it measured 2.84 CoreMark/MHz when realized with the FPGA, which is in accordance with our expectations in \S\ref{s:design}.
The core also integrates a FPU~\cite{hauser2019berkeley} that is powered but not clocked or supported at the instruction level, which is {\em for silicon area and static power accounting} as it has nothing to do with the integer pipeline.
As shown in Table~\ref{tbl:ourcpu}, its kernel-level interrupt latency is way less than Cortex-M3~\cite{arm_cortex_m3_trm} from the ARM DesignStart package, and is on par with IBEX~\cite{ibex}, which is another open-source RISC-V core.

\head{Discussion.}
Our core rivals existing cores in both interrupt latency and floorplan area.
CoreMark shows multipliers improve performance significantly at modest area expenses, while branch prediction gives a little boost for a little area.

\subsection{Control and Status Register Scheme}
We define several CSRs to support our user-level interrupt extension.
The {\tt muictl} acts as the master control: bits [31:2] hold the word-aligned base address of the IID when implemented as an in-memory table, bit 1 indicates hardware support for the extension, and bit 0 globally enables or disables the feature.
When disabled, the processor operates as a standard RISC-V core, preserving full backward compatibility.
The {\tt muistk} stores the base address of a dedicated stack onto which the hardware automatically saves general-purpose and critical control registers upon a user-level interrupt.
The {\tt muiepc}, similar to standard RISC-V {\tt mepc}, preserves the program counter of the interrupted instruction.
Upon user-level interrupt nesting, {\tt muiepc} is also pushed to the stack with the general-purpose registers for nested resumption.

When IID is implemented with CAM, each CAM entry contains three associated registers called {\tt iidnumX}, {\tt iidpmpX} and {\tt iidtimX}, where X denotes the index.
Among them, the {\tt iidnumX} holds the interrupt number for parallel matching against incoming interrupt requests.
Upon a match, the {\tt iidpmpX} provides the pointer to the corresponding PMP configuration in RAM, while the {\tt iidtimX} provides the pointer to the associated execution budget entry in RAM.

\subsection{V1: Minimal Complexity}
\label{ss:t1}
We begin our exploration with the most straightforward variant {\bf (G4)}, aiming at minimizing area and power overheads.
This variant avoids adding memory interfaces or register banks to the base core, instead performing all interrupt-related table lookups and context dumps exclusively via main SRAM.
Accordingly, we organize the IID as a linear table, and co-locate the IID table, PMP table, execution budget table, and register context stack in main SRAM.
On preemption, kernel-managed PMP context is also spilled to main SRAM.

This yields a core area of 0.1075 mm$^2$ (+16\%) and an interrupt latency of 44 cycles, with static power of 2.36 mW and dynamic power of 31.45 $\mu$W/MHz.
As shown in Figure~\ref{fig:c1var}, when kernel-managed PMP context is instead stored in dedicated shadow registers, the latency drops to 38 cycles, with a core area of 0.1105 mm$^2$ (+19\%), static power of 2.44 mW, and dynamic power of 31.75 $\mu$W/MHz.

\head{Discussion.}
While 38 cycles represents a improvement over kernel mediations that cost hundreds of cycles, this remains substantially higher than kernel-level interrupts.
Nevertheless, this minimal variant serves as a compelling choice for designs that require strict drop-in compatibility with existing bus architectures and prohibit additional memory ports.
We draw three key observations:
\begin{inparaenum}[(1)]
\item shadowing kernel-managed PMP context incurs only minimal area overhead,
\item register context stacking is the longest memory operation,
\item {\em serializing all memory accesses onto a single bus leads to severe contention}, the most critical insight guiding subsequent optimizations.
\end{inparaenum}
Since shadowing kernel-managed PMP context increases core area by less than 0.003 mm² (+3\%), we adopt this optimization in all implementations and omit further discussions of this facet.

\subsection{V2: Stack in TCM}
\label{ss:t2}
To mitigate bus contention, we introduce an additional TCM port dedicated to the register context stacking, which avoids bus conflicts when this is done in parallel with other bus activities.
While this may compromise bus interface compatibility, many mainstream processors---including the ARM Cortex-M and Cortex-R---integrate TCMs natively, where leveraging them is free.

This implementation reduces the latency to 29 cycles as shown in Figure~\ref{fig:c2var}, with a core area of 0.1058 mm$^2$, static power of 2.30 mW, and dynamic power of 32.60 $\mu$W/MHz.
Note that the area is {\em even smaller} than that of \S\ref{ss:t1} due to simplification of the SRAM-side port multiplexer.

\head{Discussion.}
Comparing this implementation with {\bf V1} (\S\ref{ss:t1}), we draw two key conclusions:
\begin{inparaenum}[(1)]
\item adding dedicated TCM ports introduces external memory interface complexity but incurs negligible area and power overhead, and
\item {\em register context stacking remains dominant even with a dedicated memory port}, which renders further port dedication to the PMP table and budget table ineffective at this stage.
\end{inparaenum}

\subsection{V3: V2 plus Banked Registers}
\label{ss:t3}
To fully eliminate the register context stacking bottleneck, we employ additional register banks to enable zero-latency context switching, and keep the dedicated stack TCM port in {\bf V2} (\S\ref{ss:t2}) to support unbounded nesting once the banks are expended.
The timing diagram is similar to that of {\bf V2}'s Figure~\ref{fig:c2var}, except that the instruction fetching now immediately follows budget updating. 

Adding one extra register bank reduces interrupt latency to 17 cycles, with a core area of 0.1257 mm$^2$ (+35\%), static power of 2.73 mW, and dynamic power of 34.70 $\mu$W/MHz.
Further increasing the number of extra banks to 2 or 3 raises the core area to 0.1530 mm$^2$ (+64\%) and 0.1683 mm$^2$ (+81\%), respectively, representing a steep escalation.

\head{Discussion.}
This implementation confirms that register banking effectively {\em removes the context stacking bottleneck} but incurs significant area and static power overhead when compared with {\bf V2} (\S\ref{ss:t2}).
Moreover, to maintain zero-latency banking at deep nesting levels, the area penalty is prohibitive in cases where cost or power efficiency is a must.
Nevertheless, it matches the interrupt latency of Cortex-R4, a popular automotive and medical processor.

\subsection{V4: V3 plus Tables in TCM}
\label{ss:t4}
With the stacking bottleneck fully resolved, the critical path now shifts to contention between the PMP table and the execution budget table, which are both in SRAM.
We therefore choose to add one more TCM port and move the PMP table to its dedicated TCM.
The timing diagram still resembles that of {\bf V2}'s Figure~\ref{fig:c2var}, except that the instruction fetching now immediately follows PMP updating, and the PMP updating happens in parallel to budget updating. 

This refinement reduces interrupt latency to 14 cycles, with a core area of 0.1269 mm$^2$ (+36\%), static power of 2.74 mW, and dynamic power of 34.50 $\mu$W/MHz.

\head{Discussion.}
From a practical engineering standpoint, this result is already compelling: it roughly matches the kernel-level interrupt latency of the Cortex-M0, which is prevalent in Unmanned Aerial Vehicles (UAVs).
The only bottleneck that remains is the IID, whose lookup {\em must be performed} before all other activities could be launched in parallel.

\begin{figure}[tb]
  \centering
  \includegraphics[width=\linewidth, trim=0.35cm 0.10cm 0.48cm 0cm, clip, keepaspectratio]{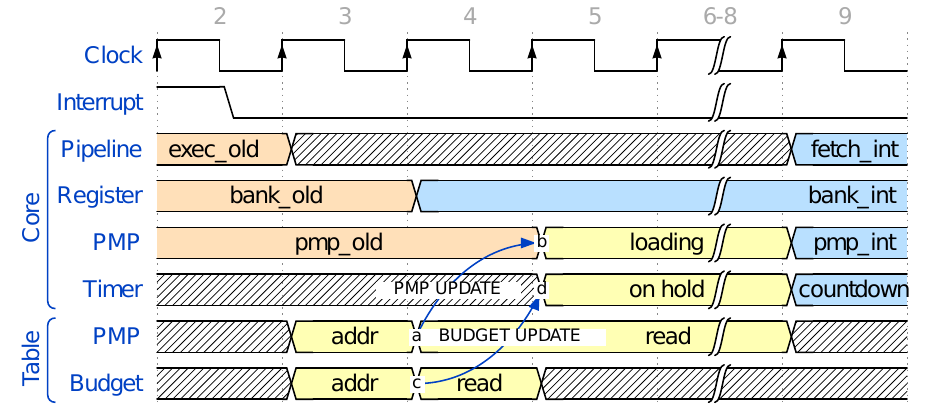}
  \caption{\small Timing diagram of the V5 variant upon user-level interrupt activation, with a latency of 11 cycles. The commentary is the same as that of Figure~\ref{fig:c1var}, except that the IID consulting is merged with interrupt controller logic and cost no extra cycles, while the register stacking is eliminated by banking; hence they are not shown.}
  \label{fig:c5var}
\end{figure}

\subsection{V5: V4 plus IID as CAM}
\label{ss:t5}
Based on {\bf V4} (\S\ref{ss:t4}), we implement the IID as CAM to embed the lookup operation directly into the interrupt acknowledge cycle.
This CAM is configured to store up to 16 entries; upon a lookup, the first matching entry outputs the corresponding PMP table entry address and budget table entry address.

As shown in Figure~\ref{fig:c5var}, this final optimization design reduces interrupt latency to 11 cycles, with a core area of 0.1372 mm$^2$ (+47\%), static power of 2.89 mW and dynamic power of 33.75 $\mu$W/MHz.
Further increasing the number of entries to 32 or 48 raises the core area to 0.1514 mm$^2$(+63\%) and 0.1605 mm$^2$(+72\%), which is even steeper than register banking.

\head{Discussion.}
This optimization delivers near-minimum latency at the expense of substantially increased area, and is already faster than the Cortex-M3, which is widely deployed in Industry Control Systems (ICS).
While the IID CAM optimization is logically independent of other techniques, it only yields a 3-cycle latency reduction with an additional 12\% area overhead, making it {\em attractive only when combined with other complementary optimizations}.
For comparison, register banking---the most costly optimization---delivers a 9-cycle reduction at a cost of 16\% area, which is a 2.25$\times$ more favorable trade-off.
Notably, this variant exhibits lower dynamic power than both V4 and V3, which stems from the decoupling of the IIDU module from the bus.

\subsection{Implementation Summary}
\begin{table}[H]
\centering
\scriptsize 
\caption{\small Characteristics of different processor configurations, including those with different extension variants. Floorplan area is in mm$^2$, interrupt latency is in CPU cycles, and power is in static (mW) / dynamic ($\mu$W/MHz) format. ``Base'' refers to our RV32IM core, and ``Kernel'' refers to the microkernel software path in~\cite{pan18mpu}.}
\begin{tabular}{|c|c|c|c|}
  \hline Configuration    & Interrupt latency             & Floorplan area             & Power \\
  \hline {\bf Base}       & {\bf 5 (kernel-level)}        & {\bf 0.0931 (100\%)}       & {\bf 2.07/30.48} \\
  \hline {\bf Base+V1}    & {\bf 38}                      & {\bf 0.1105 (119\%)}       & {\bf 2.44/31.75} \\
  \hline {\bf Base+V2}    & {\bf 29 (-9)}                 & {\bf 0.1058 (114\%)}       & {\bf 2.30/32.60} \\
  \hline Base+V3          & 17 (-21)                      & 0.1257 (135\%)             & 2.73/33.85 \\
  \hline Base+V4          & 14 (-24)                       & 0.1269 (136\%)             & 2.74/35.10 \\
  \hline {\bf Base+V5}    & {\bf 11 (-27)}                 & {\bf 0.1372 (147\%)}       & {\bf 2.89/33.75} \\
  \hline Kernel          & 634                           & N/A                        & N/A \\
  \hline
\end{tabular}
\label{tbl:selected}
\end{table}

Table~\ref{tbl:selected} summarizes the latency, area and power characteristics of all implementations.
Each resolves one critical performance bottleneck at a time, with successively diminishing returns.
Considering the diverse requirements of real-world applications, we highlight three key configurations for in-depth evaluation:
\begin{itemize}[topsep=1pt,itemsep=0pt,parsep=1pt,itemindent=0pt,leftmargin=0.3in]
\item[{\bf V1:}] {\bf IID as table, register context on stack, no extra TCM ports.}
  This variant minimizes area and requires no modifications to the memory interface, but incurs the longest latency. It is ideal for cores demanding strict drop-in compatibility with existing designs.
\item[{\bf V2:}] {\bf IID as table, register context on stack, dedicated TCM for stack.}
  This variant reduces interrupt latency by adding a single dedicated TCM port.
  Although it complicates the memory interface, it introduces negligible area and power overhead.
  It is well-suited for cores with native TCM support or strict area constraints.
\item[{\bf V5:}] {\bf IID as CAM, banked register context, dedicated TCM for protection tables, dedicated TCM for register spills.}
  This variant achieves the lowest interrupt latency, but incurs the highest area and power overhead, plus two additional TCM ports.
  It is exclusively for applications requiring absolute minimum latency.
\end{itemize}

\section{Evaluation}
\label{s:eval}
In our evaluation, we examine the three representative implementations listed in \S\ref{s:impl}, and aim to answer the following questions that directly corresponds to our goals listed in \S\ref{s:design}:
\begin{itemize}[topsep=1pt,itemsep=0pt,parsep=1pt,itemindent=0pt,leftmargin=0.3in]
\item[{\bf Q1:}] {\bf Does the extension exhibit consistent low interrupt latencies regardless of the current process, when compared with the state of the art (G1)?}
\item[{\bf Q2:}] {\bf Does the extension terminate the execution of interrupt handler after it breaks spatial (G2) or temporal (G3) isolation?}
\item[{\bf Q3:}] {\bf Is the area cost acceptable, given the fact that CPU is only a portion of the entire chip (G4)?}
\item[{\bf Q4:}] {\bf What changes to kernel and application is needed to leverage this extension (G5)?}
\end{itemize}

\begin{figure}[htb]
  \centering
  \includegraphics[width=0.5\textwidth, trim=0 8 0 0, clip]{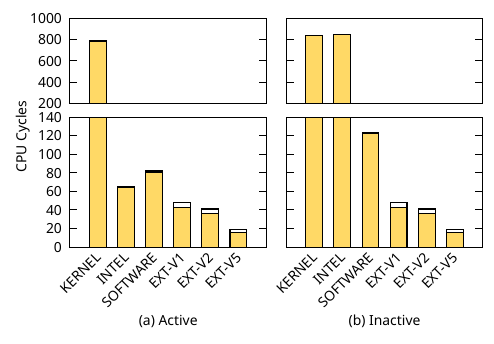}
  \caption{\small Raw interrupt latency, in CPU cycles. Lower is better.}
  \label{fig:latency}
\end{figure}

\subsection{Evaluation Setup}
We implement the aforementioned RISC-V processors respectively with three key variants on a Xilinx XC7K410T FPGA platform.
The system-on-chip (SoC) runs at 50 MHz, a typical for such processors, and integrates 128 KiB of on-chip RAM and 256 KiB of Flash, with the latter emulated using RAM.
All test binaries are compiled using GCC 15.2.0 with the -O3 optimization level to reflect real-world deployment performance.
Each measurement is repeated 10,000 times to ensure statistically robust results.

In all bar figures, average and maximum values are presented as stacked bars: {\em the darker foreground bar denotes the average-case, while the lighter upper bar represents the worst-case}.
In all line plots, data points are measured at {\em 10\%, 20\%, 50\%, and 100\%} of each scheme's maximum application performance; intermediate values in discussion are obtained via linear interpolation.

To evaluate our proposed extension, we compare its three variants against three representative baselines:
\begin{inparaenum}[(1)]
\item {\tt KERNEL}, the conventional user-level interrupt mechanism relying entirely on kernel mediation~\cite{pan2025fvm},
\item {\tt INTEL}, the Intel- or RISC-V-N-style scheme that bypasses the kernel only for active target processes;
we emulate it by handling interrupts directly in kernel mode if the target process is active, while falling back to {\tt KERNEL} otherwise~\cite{intel2025swmanualv3,rv2025next}, and
\item {\tt SOFTWARE}, a lightweight software-only scheme without hardware privilege mode transitions: it uses minimal prologues to save context and reconfigure PMP---the PMP is more efficient than software bounds checking on-the-fly~\cite{amit17tockos,peach20ewasm}---before dispatching to the handler directly in kernel mode, which serves as the near-optimal latency achievable for pure software interrupt handling~\cite{daniel04safersloth,ma2025hopter}.
\end{inparaenum}

\subsection{Raw Interrupt Latency ({\bf Q1}).}
To start, we measure the latency from interrupt signal pin assertion to the execution of the first instruction in the user-level handler.
To produce the interrupts, an auto-reloading decrementing timer is configured to fire every 4,000 cycles, far longer than the latency of all methods.
The latency is computed as the difference between the fixed overflow period and the remaining timer count captured upon handler entry, minus the timer read which is 3 cycles.

As shown in Figure~\ref{fig:latency}, we measure the latency under two scenarios: (a) when the target process is actively running, and (b) when it is inactive and has been switched out.

\head{Discussion.}
All measured latencies are highly deterministic, with maximum values closely aligning with average values across all schemes when the target process is known to be active or inactive.
{\tt KERNEL} exhibits consistently poor performance, with latency exceeding 800 cycles even in the best case.
Since its critical path is fully dominated by kernel mediation, latency varies little between active and inactive conditions, but its extremely high absolute latency renders it unsuitable for low-latency real-time systems.
{\tt INTEL} shows extreme performance variability: while it achieves moderate latency for active processes by bypassing the kernel, it incurs massive overhead in the inactive state as it falls back to the full {\tt KERNEL} path.
{\tt SOFTWARE} hits a fundamental latency floor, with around 120 cycles of latency when inactive, roughly 30 cycles of which stem from PMP configuration.
In contrast, our extensions maintain latencies below 50 cycles across conditions, with near-zero variance between active and inactive scenarios, a feat unattainable by software or state-dependent hardware schemes.
Notably, {\tt V5} achieves under 20 cycles, outperforming {\tt KERNEL} by over 40$\times$ and {\tt SOFTWARE} by roughly 6$\times$.
Even {\tt V1} and {\tt V2} outperforms {\tt SOFTWARE} by more than 3$\times$ when the target process is inactive.

\subsection{Isolation Effectiveness ({\bf Q2}).}
To evaluate the security of our isolation mechanism, we set up malicious interrupt handlers to perform unauthorized memory accesses (security and integrity) or execute infinite loops (availability).
We test three critical preemption scenarios:
\begin{inparaenum}[(1)]
\item the malicious handler preempts a user-level thread,
\item the malicious handler preempts the kernel, and
\item the malicious handler preempts another user-level interrupt handler.
\end{inparaenum}

\head{Discussion.}
Our experimental results demonstrate that all extension variants enforce strict temporal and spatial protection domains across all preemption scenarios.
Any malicious handler attempting to violate isolation guarantees is immediately and forcibly terminated, with control returned to the preempted context, safeguarding system security, integrity and availability.

\subsection{Latency-sensitive Application ({\bf Q1})}

\begin{figure}[tb]
  \centering
  \includegraphics[width=0.5\textwidth, trim = 0 8 0 0, clip]{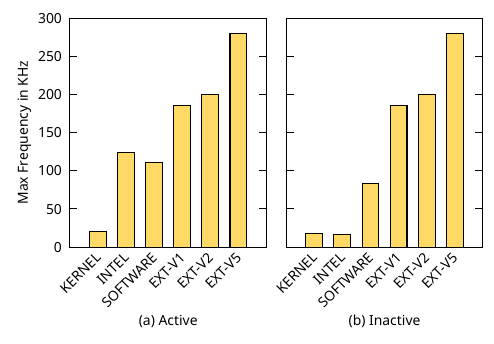}
  \caption{\small Maximum achievable PTO frequency. Higher is better.}
  \label{fig:pto}
\end{figure}

\begin{figure}[tb]
  \centering
  \includegraphics[width=0.5\textwidth, trim= 0 6.2 0 0, clip]{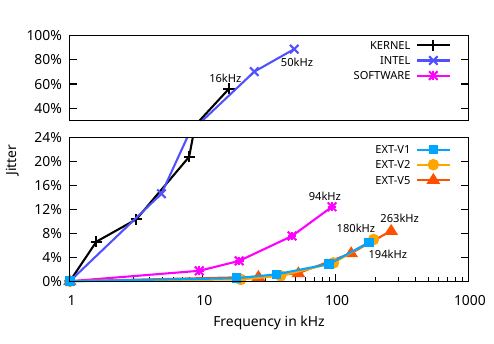}
  \caption{\small Normalized jitter as a function of PTO frequency. Lower is better.}
  \label{fig:jitter}
\end{figure}

To demonstrate the effectiveness of our extensions in latency- and jitter-sensitive scenarios, we apply them to the Pulse Train Output (PTO) application~\cite{s71200manual}, which is a core function in Programmable Logic Controllers (PLCs), robotic systems, and UAVs.
It requires the system to generate precisely timed pulse sequences to drive actuators and execute motion commands.
Typically, a background task prepares the pulse train specification including initial pulse period and per-pulse period delta.
A foreground interrupt handler then computes the required pulse intervals on-the-fly and configures the timer for next timeout accordingly, so that the output pin toggles timely upon interrupt firing.
For this evaluation, we use a square-wave pulse train to measure two critical metrics:
\begin{inparaenum}[(1)]
\item the maximum sustainable PTO frequency, correlating with speed or resolution, and
\item normalized pulse waveform jitter across frequencies, correlating with accuracy.
\end{inparaenum}

Figure~\ref{fig:pto} reports the maximum sustainable PTO frequencies across all schemes, measured under both active and inactive target process states.
To capture the dynamic and unpredictable workloads typical of real-world robotic and industrial control systems, Figure~\ref{fig:jitter} characterizes timing jitter as a function of PTO frequency under mixed process conditions, where the target process and an unrelated background process rapidly interleave.

\head{Frequency Discussion.}
The {\tt KERNEL} baseline yields the lowest PTO frequencies, as its heavy kernel mediation makes it unsuitable for fast actuation which requires at least 20 kHz~\cite{s71200manual}.
The {\tt INTEL} performs adequately when the target process is active, however it degrades to {\tt KERNEL} otherwise, while the {\tt SOFTWARE} provides a middle-ground.
In contrast, our extensions are nearly unaffected by process state, consistently outperforming all baselines and achieving peak PTO frequencies up to near 200 kHz.

\head{Jitter Discussion.}
Under the mixed process scenario, {\tt KERNEL} and {\tt INTEL} both exhibit catastrophic degradation, exceeding 60\% jitter at just 16 kHz, rendering them unusable for any accurate actuator control task.
{\tt SOFTWARE} on the other hand has better jitter, but remains unacceptable for precision applications that require less than 10\% jitter at 100 kHz~\cite{s71200manual}.
Instead, our extensions maintain less than 5\% jitter at 100 kHz, with the {\tt V5} achieving 8\% jitter at over 250 kHz, satisfying the most demanding applications.
At 10kHz which is a common frequency, all our extensions exhibit less than 0.5\% jitter, which is 60$\times$ better than kernel-mediated baselines.

\subsection{Task-colocation Application ({\bf Q1})}

\begin{figure}[ht]
  \centering
  \includegraphics[width=0.5\textwidth, trim= 0 10 0 0, clip]{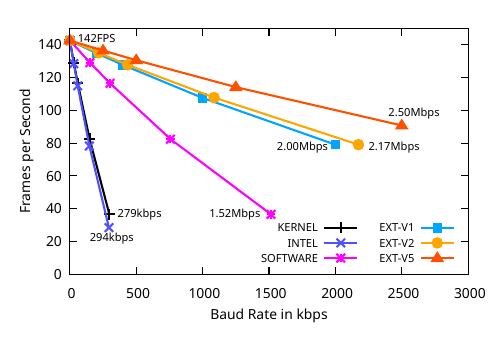}
  \caption{\small Achievable FPS of a background image recognition task as a function of foreground Modbus-RTU baud rate. Higher is better.}
  \label{fig:modbus}
\end{figure}

In domain controller consolidation scenarios~\cite{autosar2020}, foreground real-time tasks are often colocated with background best-effort workloads to cut hardware costs, a trend also seen in cloud computing~\cite{li2023learning}.
The {\em interference is mutual}: the real-time task suffers prolonged {\em worst-case} latencies preempting the continuously running background batch job, while the best-effort job suffers severe {\em average-case} performance degradation due to frequent interrupts from the communication task.

We use Modbus-RTU---a widely adopted serial protocol in industrial robots and PLCs~\cite{tamboli2015impl}---as the real-time communication workload.
It generates one interrupt per byte transmitted, resulting in high interrupt frequency that stresses the interrupt handling pipeline.
The background best-effort task is a real-world image recognition workload that repeatedly performs digit recognition on input images, a common scenario in edge vision systems.
As shown in Figure~\ref{fig:modbus}, we measure the impact of interrupt handling overhead on the background task's throughput (Frames per Second, FPS) as the Modbus baud rate increases.

\head{Discussion.}
Under the inactive process condition enforced in this experiment, {\tt INTEL} exhibits nearly identical behavior to the {\tt KERNEL} baseline, as it always falls back to the kernel-mediated slow path, since the background task remains running continuously.
While both schemes readily support 115.2 kbps, a widely used industrial baud rate, their background task performance degrades severely, cutting FPS from 142 to 85, which is a 40\% reduction.
Furthermore, neither can sustain baud rates above 300 kbps due to their excessive interrupt latency, a critical limitation for modern UAVs and industrial systems.
{\tt SOFTWARE} outperforms the kernel-mediated baselines and can operate at 1 Mbps, but at this baud rate, FPS drops to 68, representing a 52\% performance loss.
Notably, all our hardware extensions support baud rates up to 2 Mbps---twice the requirement of the most demanding applications.
At 1 Mbps, our best-performing {\tt V5} achieves 122 FPS, incurring less than 15\% loss; even the entry-level {\tt V1} sustains 107 FPS, which is a 25\% penalty.
At 115.2 kbps, all of our extensions incur at most 5 FPS penalty, which is 11$\times$ better than kernel-mediated solutions.

\subsection{Kernel Modification Impact ({\bf Q4})}
\begin{table}[h]
\scriptsize
\centering
\begin{tabular}{|c|c|c|c|}
  \hline Kernel+HAL    & V1              & V2              & V5 \\
  \hline 5856+4249     & +216 (+2.16\%)  & +216 (+2.16\%)  & +270 (+2.70\%) \\
  \hline
\end{tabular}
\caption{Modifications of microkernel source, in SLoC.}
\label{tbl:kernmod}
\end{table}

To assess the adaptation effort on existing microkernel code, we measure the Source Lines of Code (SLoC) that must be added to leverage the extensions, with the results shown in Table~\ref{tbl:kernmod}.

\head{Discussion.}
We may observe that the increased lines of code is small when compared to the entire code base, reflecting minimal adaptation effort.

\subsection{Die Area Impact ({\bf Q5})}
\begin{table}[h]
\scriptsize
\centering
\caption{\small Total die area of typical SoC. Length and width are in mm, while area are in mm$^2$. Core\% are the portion of our base core on the same assumed die area, V1\%, V2\% and V5\% is the extra area that the V1, V2 and V5 extension variant would require. The reason for a very small Core\% is because SoC integrate RAM, Flash and various analog peripherals, which takes up quite some space.}
\begin{tabular}{|c|c|c|c|c|c|}
  \hline Die        & L$\times$W = Area         & Core\%  & V1\% & V2\% & V5\% \\
  \hline STM32H743  & 5.06$\times$4.70 = 23.78  & 0.4\%   & 0.07\% & 0.06\% & 0.19\% \\
  \hline LPC55S69   & 2.90$\times$2.78 = 8.06   & 1.1\%   & 0.22\% & 0.19\% & 0.55\% \\
  \hline ESP32C3    & 2.28$\times$2.27 = 5.18   & 1.8\%   & 0.34\% & 0.30\% & 0.85\% \\
  \hline TC387      & 7.64$\times$7.39 = 56.46  & 0.2\%   & 0.03\% & 0.03\% & 0.08\% \\
  \hline
\end{tabular}
\label{tbl:die}
\end{table}

We decapped four SoC implemented using 40 nm technology (more compact than our 45 nm) and obtained their die area, to evaluate the area impact of our extension on the entire die.

\head{Discussion.}
It could be observed from Table~\ref{tbl:die} that CPU core accounts for a tiny portion of the total die area, particularly noting that our core uses 45 nm process technology.
As a result, our extensions account for less, and is negligible in some cases. Even if we assume a much smaller die such as 4 mm$^2$, the largest V5 extension would account for 2\% of extra area.

\section{Conclusions}
\label{s:conc}
This paper proposes a hardware extension that aims to achieve deterministic user-level interrupt delivery.
Through design and implementation, we confirm the implementation feasibility of the proposal and present three variants, which reduces worst-case latency by more than 50× with a 19\% increase in core area (2\% of total die area) and 4.1\% increase in dynamic power.
The extension provides a comprehensive solution for secure, reliable and deterministic cyber-physical systems.
We also note that similar extensions might be possible for MMU-based general-purpose systems as future research.



\end{document}